# Low-Latency SC Decoder Architectures for Polar Codes

Chuan Zhang, Bo Yuan, and Keshab K. Parhi, *Fellow, IEEE*

*Abstract*—Nowadays polar codes are becoming one of the most favorable capacity achieving error correction codes for their low encoding and decoding complexity. However, due to the large code length required by practical applications, the few existing successive cancellation (SC) decoder implementations still suffer from not only the high hardware cost but also the long decoding latency. This paper presents novel several approaches to design low-latency decoders for polar codes based on look-ahead techniques. Look-ahead techniques can be employed to reschedule the decoding process of polar decoder in numerous approaches. However, among those approaches, only well-arranged ones can achieve good performance in terms of both latency and hardware complexity. By revealing the recurrence property of SC decoding chart, the authors succeed in reducing the decoding latency by 50% with look-ahead techniques. With the help of VLSI-DSP design techniques such as pipelining, folding, unfolding, and parallel processing, methodologies for four different polar decoder architectures have been proposed to meet various application demands. Sub-structure sharing scheme has been adopted to design the *merged processing element* (PE) for further hardware reduction. In addition, systematic methods for construction refined pipelining decoder (2$^{nd}$ design) and the *input generating circuits* (ICG) block have been given. Detailed gate-level analysis has demonstrated that the proposed designs show latency advantages over conventional ones with similar hardware cost.

## I. INTRODUCTION

INTRODUCED by Arıkan recently [1], polar codes have shown the capabilities to achieve the symmetric capacity $I(W)$ of any given binary-input discrete memoryless channel (B-DMC) $W$. Considered as the first "low complexity" scheme which provably achieves the capacity for a fairly wide array of channels, polar codes have become one of the most favorable research topics. By recursively combining and splitting the $N$ copies of B-DMC, we obtain a second channel $W_N$, which is composed of $W_N^{(i)}$ with $1 \leq i \leq N$. Among those $N$ newly constructed channels, only those with highest capacity are used for data transmission. We refer outputs of these channels as information bits, and the set of corresponding indices as $\mathcal{A}$. Also the outputs of the other channels are denoted as frozen bits, and these channels' indices make up the set $\mathcal{A}^c$.

Although a great deal of research effort has been expended in study of polar codes, most of the research is focused on code performance rather than the design of high efficiency decoders. Shown in [1], the straightforward polar decoder implementation with successive cancellation (SC) algorithm results in the complexity of $\mathcal{O}(N\log_2 N)$. A simple implementation approach based on belief propagation (BP) algorithm was proposed by the same author in [1]-[2]. However, due to its lower complexity compared with BP algorithm, the SC approach appears more attractive for hardware designers. Therefore, a reduced SC decoder with complexity of $\mathcal{O}(N)$ was given by [3]. For the decoding of a polar code with length of $N$, totally $2(N-1)$ clock cycles are required by the decoder. And in each active stage, the highest hardware efficiency can be only 50%, which means more than half processing elements (PEs) are idle at the same time. This is because the estimation of the current bit also depends on the previous coded bit, which forces all coded bits to be output successively. In order to achieve faster decoding, the loop computation can be reformulated based on look-ahead techniques, which pre-calculate all possible outputs of the next code bit and then select the correct one with a multiplexer. However, among all possible candidates, only the one with short latency and low hardware complexity can be selected. This paper addresses one nice recursive time chart construction method which succeeds in reducing the decoding latency with look-ahead techniques in any cases. By employing design techniques such as pipelining, folding, unfolding, and parallel processing, several general design methods for polar decoders are presented accordingly. Benefitting from the efficient real FFT processor architecture in [4], the *input generating circuits* (ICG) block is proposed to can generate all selective signals on the fly. Along with the reduced-complexity *merged PE*, all gate-level design details for proposed polar decoders are well illustrated. Comparison results have shown that each design approach given in this paper is able to achieve only half decoding latency while consumes comparable hardware as the conventional ones, which is attractive for high speed real-life applications.

The remainder of this paper is organized as follows. A brief review of SC decoding algorithm and its logarithm domain variants are provided in Section II. In Section III, the conventional decoding time chart is regenerated with recurrence relationship. And the systematic algorithm to construct the look-ahead scheduling scheme is given in a recursive manner. The corresponding latency-reduced polar decoder architecture (1st decoder) and its sub-blocks, such as the *merged PE* and the ICG module is discussed in block Section IV. In Section V, based on the one given in Section IV three modified decoders are presented, which employ unfolding (2nd decoder), folding (3rd decoder), and parallel processing techniques (4th decoder), respectively. The 2nd decoder architecture is compatible with $M$ consecutive inputs processing. The 3rd one time-multiplexes all decoding operations on a single PE stage. Using the same number of PEs as the 3rd one, the 4th decoder manages to implement 2-parallel processing at the price of an additional clock cycle. It is obvious to see that compared with the 1st design, each variant improves the hardware efficiency while keeps the same low-latency advantage. The corresponding performance estimation and comparison with state-of-the-art designs are presented in Section VI. Section VII concludes the paper.

## II. REVIEW OF SC ALGORITHM AND ITS VARIANTS

In this section, we provide the preliminaries of the SC decoding algorithm. Moreover, some variants and simplified modifications of the SC algorithm are explained as well.

### A. SC Decoding Algorithm

Consider an arbitrary polar code with parameter ($N$, $K$, $\mathcal{A}$, $u_{\mathcal{A}^c}$) [1]. We denote the input vector as $u_1^N$, which consists of a random part $u_{\mathcal{A}}$ and a frozen part $u_{\mathcal{A}^c}$. The corresponding output vector through channel $W_N$ is $y_1^N$ with conditional probability $W_N(y_1^N | u_1^N)$. Define the likelihood ratio (LR) as,

$$L_N^{(i)}(y_1^N, \hat{u}_1^{i-1}) \triangleq \frac{W_N^{(i)}(y_1^N, \hat{u}_1^{i-1} | 0)}{W_N^{(i)}(y_1^N, \hat{u}_1^{i-1} | 1)}. \quad (1)$$

The a posteriori decision scheme is given as follows,

---

**A Posteriori Decision Scheme with Frozen Bits**

1: **if** $i \in \mathcal{A}^c$ **then** $\hat{u}_i = u_i$;
2: **else**
3:     **if** $L_N^{(i)}(y_1^N, \hat{u}_1^{i-1}) \geq 1$ **then** $u_i = 0$;
4:     **else** $\hat{u}_i = 1$;
5:     **endif**
6: **endif**

---

It is noted that LRs with even and odd indices can be generated by applying the recursive formulas given by Eq. (2) and (3), respectively:

$$L_N^{(2i)}(y_1^N, \hat{u}_1^{2i-1}) = [L_{N/2}^{(i)}(y_1^{N/2}, \hat{u}_{1,o}^{2i-2} \oplus u_{1,e}^{2i-2})]^{1-2\hat{u}_{2i-1}} \cdot L_{N/2}^{(i)}(y_{N/2+1}^N, u_{1,e}^{2i-2}), \quad (2)$$

$$L_N^{(2i-1)}(y_1^N, \hat{u}_1^{2i-2}) = \frac{L_{N/2}^{(i)}(y_1^{N/2}, \hat{u}_{1,o}^{2i-2} \oplus u_{1,e}^{2i-2}) L_{N/2}^{(i)}(y_{N/2+1}^N, u_{1,e}^{2i-2}) + 1}{L_{N/2}^{(i)}(y_1^{N/2}, \hat{u}_{1,o}^{2i-2} \oplus u_{1,e}^{2i-2}) + L_{N/2}^{(i)}(y_{N/2+1}^N, u_{1,e}^{2i-2})}. \quad (3)$$

Obviously, the calculation of $L_N^{(i)}(y_1^N, \hat{u}_1^{i-1})$ depends on the estimation of the previous bit, from which the SC decoding algorithm is named. For the ease of clear explanation, the decoding procedure of a polar code with $N = 8$ is illustrated in Fig. 1, where Type I and Type II PEs are in charge of computations given in Eq. (2) and (3), respectively. And the label attached to each PE indicates the index of clock cycle when the corresponding PE is activated.

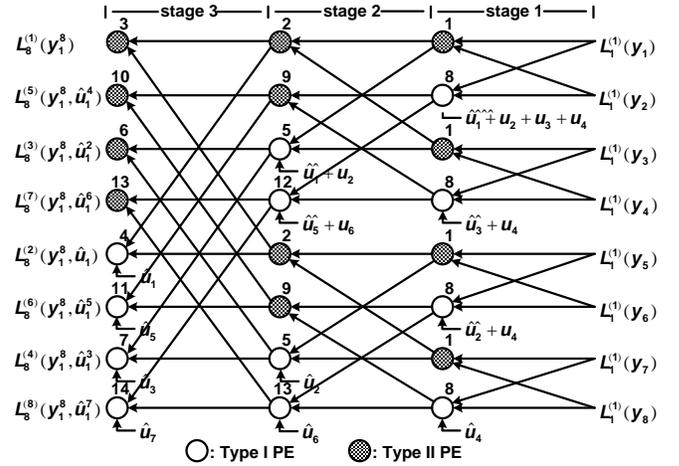

Figure 1: SC decoding process of polar codes with length $N = 8$.

### B. SC Decoding Algorithm in Logarithm Domain

For any general decoding algorithm, its variant defined in logarithm domain always has advantages in terms of hardware implementation, computational complexity, and numerical stability over the one in real domain. Therefore, similar to the approach addressed in [5], the SC algorithm dealing with logarithm-likelihood ratio (LLR) was mentioned by [3]. Eq. (4) and (5) can then be rewritten as follows:

$$\mathbb{L}_N^{(2i)}(y_1^N, \hat{u}_1^{2i-1}) = (-1)^{\hat{u}_{2i-1}} \mathbb{L}_{N/2}^{(i)}(y_1^{N/2}, \hat{u}_{1,o}^{2i-2} \oplus u_{1,e}^{2i-2}) + \mathbb{L}_{N/2}^{(i)}(y_{N/2+1}^N, u_{1,e}^{2i-2}), \quad (4)$$

$$\mathbb{L}_N^{(2i-1)}(y_1^N, \hat{u}_1^{2i-1}) = 2\operatorname{artanh}\{\tanh[\mathbb{L}_{N/2}^{(i)}(y_1^{N/2}, \hat{u}_{1,o}^{2i-2} \oplus u_{1,e}^{2i-2})] \cdot \tanh[\mathbb{L}_{N/2}^{(i)}(y_{N/2+1}^N, \hat{u}_{1,e}^{2i-2})]\}. \quad (5)$$

Here the LLRs are defined as:

$$\mathbb{L}_N^{(i)}(y_1^N, \hat{u}_1^{i-1}) \triangleq \ln L_N^{(i)}(y_1^N, u_1^{i-1}). \quad (6)$$

### C. Min-Sum SC Decoding Algorithm

In order to implement the hyperbolic tangent function and its inverse function in Eq. (5), large amount of look-up table (LUT) is required. Note that in logarithm domain, for variable $x \gg 1$, the following approximation holds:

$$\ln[\cosh(x)] \simeq |x| - \ln 2. \quad (7)$$

Consequently, Eq. (5) can be reduced to the min-sum update rule, which is LUT free:

$$\mathbb{L}_N^{(2i-1)}(y_1^N, \hat{u}_1^{2i-2})$$
$$\simeq \frac{1}{2} \cdot \left| \mathbb{L}_{N/2}^{(i)}(y_1^{N/2}, \hat{u}_{1,o}^{2i-2} \oplus u_{1,e}^{2i-2}) + \mathbb{L}_{N/2}^{(i)}(y_{N/2+1}^N, u_{1,e}^{2i-2}) \right| -$$
$$\frac{1}{2} \cdot \left| \mathbb{L}_{N/2}^{(i)}(y_1^{N/2}, \hat{u}_{1,o}^{2i-2} \oplus u_{1,e}^{2i-2}) - \mathbb{L}_{N/2}^{(i)}(y_{N/2+1}^N, u_{1,e}^{2i-2}) \right| \quad (8)$$
$$= \operatorname{sgn}[\mathbb{L}_{N/2}^{(i)}(y_1^{N/2}, \hat{u}_{1,o}^{2i-2} \oplus u_{1,e}^{2i-2})] \operatorname{sgn}[\mathbb{L}_{N/2}^{(i)}(y_{N/2+1}^N, u_{1,e}^{2i-2})] \cdot$$
$$\min[|\mathbb{L}_{N/2}^{(i)}(y_1^{N/2}, \hat{u}_{1,o}^{2i-2} \oplus u_{1,e}^{2i-2})|, |\mathbb{L}_{N/2}^{(i)}(y_{N/2+1}^N, u_{1,e}^{2i-2})|].$$

Simulation results have demonstrated that the min-sum SC decoding algorithm only suffers from little performance degradation than the optimal one while achieves a good hardware efficiency [3]. This property makes min-sum SC decoding algorithm very attractive for VLSI implementation. Therefore, in the following sections we will discuss the polar decoder design based on this sub-optimal algorithm.

However, among all those decoding algorithms pre-stated, probabilities are updated according to the same data flow illustrated in Fig. 1, which is straightforward but not efficient enough. In the next section, we present our high-performance scheme for polar decoder design. In the proposed scheme, compared to the schemes presented above, the number of clock cycles required for obtaining the estimated information bits has been reduced by 50%. Moreover, this latency-reduced scheme is suitable for any code length *N*, and can be generated in a nice recursive manner.

### III. PROPOSED LATENCY-REDUCED UPDATING SCHEME FOR POLAR DECODER DESIGN

In what follows, we present a latency-reduced updating scheme for polar decoders based on a novel look-ahead scheduling method. The proposed techniques need fewer number of clock cycles to perform the same operation as the conventional SC decoding algorithm, leading to lower decoding latency. For the straightforward SC decoding implementation of *N*-bit polar codes, totally $2(N-1)$ clock cycles are required. Careful investigation has shown that the corresponding time chart can be constructed in recursive way, which is described by the following pseudo-codes. In an effort for conciseness, in the rest of this paper, the notation $TC = \{[\mathcal{C}, TC], s\}$ is used for the left insertion of an array $\mathcal{C}$ into the previously arranged time chart TC at *Stage s*. Similarly, $TC = [TC, TC]$ simply means duplication of previous time chart.

| **Recursive Construction of Conventional Time Chart** |
| --- |
| 1: **initializtion** TC = $Null$; |
| 2: **for** $i = \log_2 N, i--,1$ **do** |
| 3: $\quad j = \log_2 N - i + 1$; |
| 4: $\quad TC = \{[j \text{ of Type I}, TC], i\}$; |
| 5: $\quad TC = [TC, TC]$; |
| 6: $\quad$ change the leftmost *j* of Type I with *j* of Type II; |
| 7: **endfor** |
| 8: **output** TC. |

Here, *i* and *j* are indices of iterative execution. "*j* of Type I" is the short for *j* copy (or copies) of Type I PE(s).

Basically, *Stage i* will be activated $2^i$ times during the whole decoding process. Therefore, the total number of clock cycles required can be calculated as follows:

$$2\sum_{i=0}^{\log_2 N - 1} 2^i = 2 \cdot \frac{(2^{\log_2 N} - 1)}{2 - 1} = 2(N-1), \quad (9)$$

which matches the general conclusion given before. According to Fig. 1, it can be observed that each PE is activated only once during the entire decoding process. In one specific clock cycle only single type of PE is active. The output LLRs will be generated the same clock cycle in which the $\log_2 N$-th stage is active. For ease of clear explanation, the polar decoding process shown in Fig. 1 is employed for detailed illustration. Since block length $N = 8$, according to Eq. (9) totally 14 clock cycles are required to finish the whole decoding process. The corresponding decoding time chart is given in Fig. 2 (a).

However, this conventional decoding approach is not suitable for practical applications for the following two reasons. First, in order to achieve required decoding performance, the code length *N* is usually set to be as large as $2^{10}$-$2^{20}$. An immediate consequence is the latency of $2(N-1)$ clock cycles is too large. Second, according to Fig. 2 (a), it is apparent that during the whole decoding process the highest hardware utilization in a specific clock cycle is only 50% (*Clock cycle* 1). As the stage index increases, the hardware efficiency will go down as low as 12.5%. For general case with code length of *N*, the minimum hardware efficiency for active stage is $1/N$ (*Clock cycle* $\log_2 N$). Since only polar codes with code length greater or equal than $2^{10}$ can achieve a good performance that approaching the channel capacity, the straightforward implementation in Fig. 1 becomes impractical for real-life applications because the lowest hardware usage can be around $2^{-10}$. Even for the pipelined tree architecture proposed by [3] in Fig. 3, the highest utilization is only 50% as well, which means half PEs are in idle state during each clock cycle.

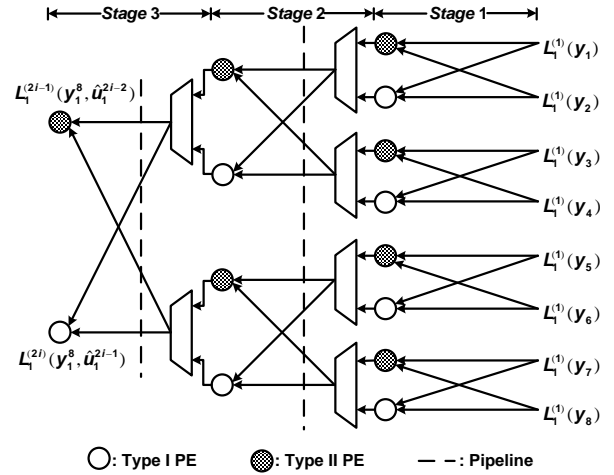

Figure 3: Pipelined decoder architectures of polar codes with length $N = 8$.

This dilemma is introduced by the bottleneck of sequential decoding property of SC algorithm. In order to address this issue properly, the computation loop can be re-scheduled with look-ahead techniques. However, in order to achieve the goal of low latency and high performance, the new decoding schedule needs to be carefully chosen. It is noted that if both the two LLR inputs for Eq. (5) are available, there are only two possible outputs. Therefore, for any Type I PE, given two deterministic LLR inputs, the look-ahead scheme only needs to pre-compute two output candidates. The correct output can be selected by a multiplexer thereafter. For the instance shown in Fig. 1, all possible outputs of Type I PEs labeled by 8 in *Stage* 1 can be pre-calculated in *Clock cycle* 1. In other words, for *Stage* 1 the required computation in *Clock cycle* 8 can be incorporated into

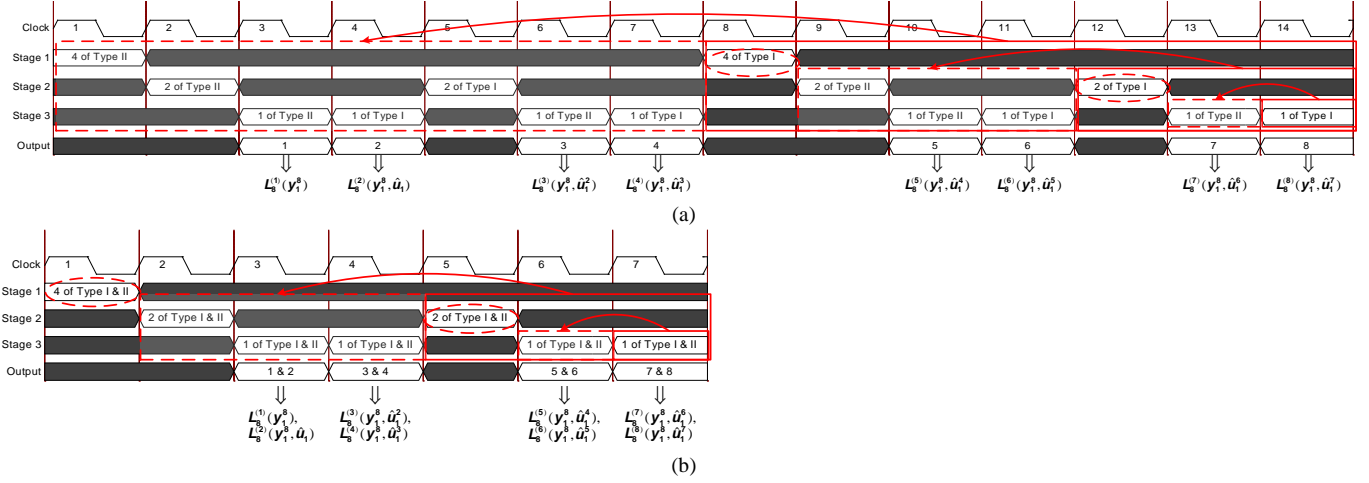

Figure 2: Conventional and look-ahead decoding time charts for polar codes with $N = 8$.

*Clock cycle* 1. In the similar way, for *Stage* 2 computation in *Clock cycle* 5 and 12 can be taken care of in *Clock cycle* 2 and 9, respectively. Calculation in *Clock cycle* 4, 7, 11, and 14 can be re-scheduled into *Clock cycle* 3, 6, 10, and 13 for *Stage* 3. As a result, only half clock cycles are required to implement the same decoding task with the help of the proposed look-ahead schedule. In general case, only the PE's with two deterministic LLR inputs can be activated in a certain clock cycle. If the corresponding PE is categorized into Type I, both two possible results are pre-calculated. Otherwise, if the corresponding PE falls into the category of Type II, the unique output is derived directly and then propagated to the next stage. For the 8-bit polar decoder example, all PEs at *Stage* 1 are activated during *Clock cycle* 1 because both deterministic LLR inputs for each PE are guaranteed by channel outputs. However, in *Clock cycle* 2, only PEs labeled with 2 and 5 can be activated, because they are the only ones with deterministic LLR inputs. For PEs with labels of 9 and 12, their LLR inputs are generated by Type I PEs in *Stage* 1, which have two possible values at this moment. In order to avoid error propagation caused by pre-computing to the next stage, those PEs stay idle during *Clock cycle* 2. Similar schemes apply to further decoding processes. Therefore, the well scheduled look-ahead decoding procedure is obtained by folding the straightforward time chart into half, which is illustrated by Fig. 2 (b). It is clear that in order to decode polar codes with length *N*, the required number of clock cycles can be halved to *N*-1. The time chart construction of the proposed new scheme is given by the following pseudo codes.

As indicated in *Step* 4 of the given construction method, benefit from the look-ahead techniques both types of PEs can work simultaneously in the same clock cycle, which not only shortens the decoding latency by 50% but also improves the hardware efficiency twice. Moreover, the proposed approach succeed in giving the construction method in a recursive way. For clear understanding of the *Russian Doll*-like relationship between stages, both conventional and look-ahead construction processes have been pointed out with arrows.

**Recursive Construction of Look-Ahead Time Chart**
1: initializtion TC=$\mathcal{N}ull$;
2: **for** $i = \log_2 N, i--, 1$ **do**
3:     $j = \log_2 N - i + 1$;
4:     TC = $\{[j \text{ of Type I \& II, TC}], i\}$;
5:     **if** $i = 1$ **then**
6:         **break**;
7:     **endif**
8:     TC = [TC, TC];
9: **endfor**
10: **output** TC.

IV. ARCHITECTURES FOR PROPOSED LOOK-AHEAD DECODER

In this section, we present the detailed hardware architectures of the proposed latency-reduced SC polar decoder using look-ahead techniques. It is known that the operations in SC decoding belong to two categories, which are executed by Type I and Type II PEs, respectively. In other words, implementing the decoder in hardware comes down to the design of two elementary elements. According to the Min-Sum SC decoding algorithm, both elementary elements are associated with addition and comparison operations, which provides us with facilities to carry out further optimization using sub-structure sharing techniques. Low-complexity structures for both two types of PEs are proposed in the rest of this section. Moreover, an optimized architecture for the *input generating circuit* (IGC) is derived as well.

*A. Design of Type I PE*

According to the look-ahead scheme, Type I PE is in charge of pre-computing two possible outputs. In order to incorporate both capabilities of adding or subtracting operands together, an adder-subtractor architecture is employed by Type I PE. For the sake of clarity, it is supposed that *q*-bit quantization is adopted here.

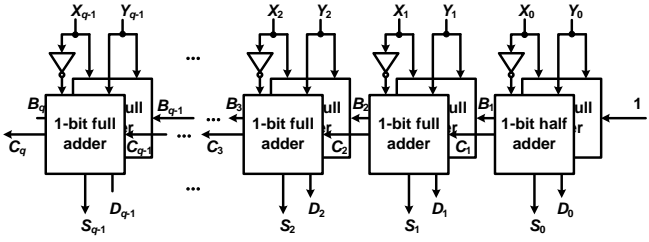

Figure 4: $q$-bit adder-subtractor architecture.

As shown in Fig.4, in order to avoid large processing latency, parallel implementation is employed here. However, simple duplication results in penalty of doubling both area and power. For a $q$-bit adder-subtractor, totally $2q$-1 1-bit full adder and one 1-bit half adder are required. In order to implement Type-I PE more effectively, the novel architecture of adder-subtractor is proposed in this section. Rather than implementing Type-I PE with two's complement approach, the original carry-borrow idea is employed here. Suppose $X$ and $Y$ are the two operands, and $Z_{in}$ is the carried-in or borrowed-from bit. For the full adder the two outputs of summation and carry-out are represented by $S$ and $C_{out}$, respectively. In similar way, the difference and borrow-out produced by the full subtructor are denoted with $D$ and $B_{out}$. Therefore, the truth table for the full adder and subtractor is given below.

TABLE I TRUTH TABLE OF BOTH FULL ADDER AND SUBTRACTOR

| Inputs | | | Outputs | | | |
|---|---|---|---|---|---|---|
| | | | Adder | | Subtructor | |
| $X$ | $Y$ | $Z_{in}$ | $S$ | $C_{out}$ | $D$ | $B_{out}$ |
| 0 | 0 | 0 | 0 | 0 | 0 | 0 |
| 0 | 0 | 1 | 1 | 0 | 1 | 1 |
| 0 | 1 | 0 | 1 | 0 | 1 | 1 |
| 0 | 1 | 1 | 0 | 1 | 0 | 1 |
| 1 | 0 | 0 | 1 | 0 | 1 | 0 |
| 1 | 0 | 1 | 0 | 1 | 0 | 0 |
| 1 | 1 | 0 | 0 | 1 | 0 | 0 |
| 1 | 1 | 1 | 1 | 1 | 1 | 1 |

From Table I one can draw the *Karnaugh* map for all outputs based on which the logic equations are derived as follows:

$$S = X \oplus Y \oplus Z_{in}; \quad (10)$$

$$C_{out} = X \cdot Y + (X \oplus Y) \cdot Z_{in}; \quad (11)$$

$$D = X \oplus Y \oplus Z_{in}; \quad (12)$$

$$B_{out} = \overline{X} \cdot Y + \overline{X \oplus Y} \cdot Z_{in}. \quad (13)$$

It is obvious that outputs $S$ and $D$ are actually the same. And easy to notice that $\overline{X} \cdot Y$ is an intermediate term of $X \oplus Y$. Similarly, $(X \oplus Y) \cdot Z_{in}$ can be treated as a byproduct of term $X \oplus Y \oplus Z_{in}$ as well. Since both outputs $C_{out}$ and $B_{out}$ can be calculated by recursively employing the AND or AND-NOT operation twice, they can be obtained simultaneously with other two outputs. The resulted gate-sharing techniques can not only implement parallel processing but also reduce the hardware consumption. The proposed gate-level structures of 1-bit full and half adder-subtractor are depicted in Fig. 5 (a) and (b) respectively, where the carry-in bit $C_{in}$ and borrow-out bit $B_{in}$ are supposed to be different rather than in the unique form of $Z_{in}$. For the sake of easy estimation and comparison, all the hardware complexities are converted in the form of equivalent XOR gate number. According to Fig. 5, the complexities of 1-bit full and half adder-subtractor equal to 4 XOR gates and 1 XOR gate, respectively. Compared with the straightforward ones, the total savings of the proposed approaches are 43% and 50%, respectively. Composed of $q$-1 1-bit full adder-subtractor and a 1-bit half adder-subtractor, the general $q$-bit adder-subtractor architecture with the given design method requires only less than 57% hardware compared with the conventional one while achieves exactly the same performances.

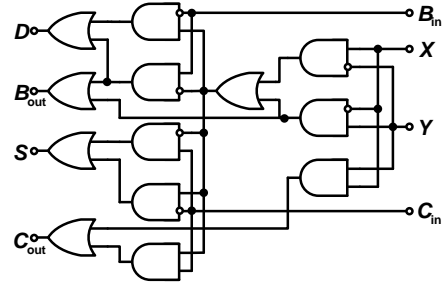

(a) 1-bit full adder-subtractor.

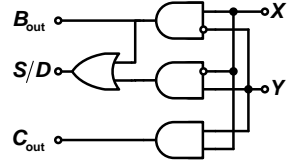

(b) 1-bit half adder-subtractor.

Figure 5: Proposed 1-bit adder-subtractor architectures.

Without being misunderstood, it is agreed that hereinafter the newly proposed Type I PE illustrated in Fig. 6 is still referred to as "Type I PE". The Type I PE with the conventional architectures will not be employed or discussed any more.

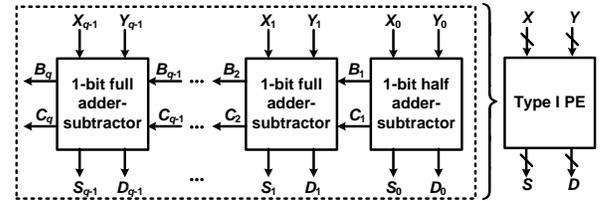

Figure 6: Proposed Type I PE architectures.

### B. Design of Type II PE

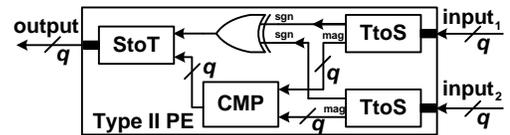

input$_1$: $\mathbb{L}_{N/2}^{(i)}(\mathbf{y}_{N/2+1}^{N}, \hat{\mathbf{u}}_{1,e}^{2i-2})$;
input$_2$: $\mathbb{L}_{N/2}^{(i)}(\mathbf{y}_{1}^{N/2}, \hat{\mathbf{u}}_{1,o}^{2i-2} \oplus \mathbf{u}_{1,e}^{2i-2})$;
output: $\mathbb{L}_{N}^{(2i-1)}(\mathbf{y}_{1}^{N}, \hat{\mathbf{u}}_{1}^{2i-2})$.

Figure 7: Proposed architectures of Type II PE.

Instead of implementing *tanh* and *artanh* functions, Type II PE which employs the min-sum algorithm is shown in Fig.7. *TtoS* block will perform the conversion from two's complement representation to sign-magnitude representation. *StoT* block will perform the reverse conversion. The *TtoS* block is illustrated in Fig. 8. In order to avoid the overflow situation, a sign extension operation is required as well.

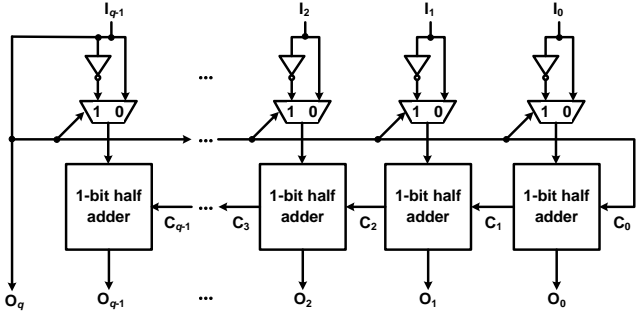

Figure 8: Proposed structure of *TtoS* block.

The *StoT* block is similar to the *TtoS* block. The only difference is that a sign compression operation is needed to make the output data in the form of the $q$-bit quantization.

### C. Sub-structure Sharing of Type I and Type II PEs

Since the comparator in Type II PE is actually a $q$-bit subtractor, which is also employed by Type I PE, it is possible to incorporate both Type I and Type II PEs together using the sub-structure sharing scheme. The detailed structure is illustrated as follows:

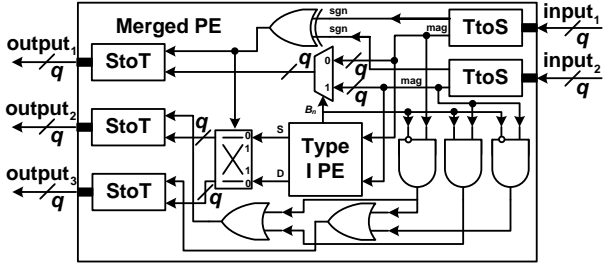

**input$_1$:** $\mathbb{L}_{N/2}^{(i)}(y_{N/2+1}^N, \hat{u}_{1,e}^{2i-2})$;  **output$_2$:** $\mathbb{L}_N^{(2i-1)}(y_1^N, \hat{u}_1^{2i-2})$, $u_{2i-1} = 0$;
**input$_2$:** $\mathbb{L}_{N/2}^{(i)}(y_1^{N/2}, \hat{u}_{1,o}^{2i-2} \oplus \hat{u}_{1,e}^{2i-2})$;  **output$_3$:** $\mathbb{L}_N^{(2i-1)}(y_1^N, \hat{u}_1^{2i-2})$, $u_{2i-1} = 1$.
**output$_1$:** $\mathbb{L}_N^{(2i-1)}(y_1^N, \hat{u}_1^{2i-2})$;

Figure 9: Proposed structure of the Merged PE.

In the Merged PE, the comparison operation is carried out by the Type I PE illustrated Fig. 6. Two more *StoT* blocks as well as additional control logic are required here. According to Fig. 9, for $q$-bit quantization scheme totally $2q-3$ XOR gates can be saved by the proposed sub-structure sharing approach. As mentioned previously, usually code length $N$ over $2^{10}$ is a must for polar codes to achieve required decoding performances in practical applications. For conventional pipelined tree polar decoder architectures, $N$-1 Merged PEs can be employed instead of $N$-1 Type I PEs and Type II PEs, respectively. As a consequence, the resulted hardware saving could be around $2^{10} \cdot (2q-3)$ XOR gates.

### D. Input Generating Circuit for Type I PEs

As indicated in Eq. (5), except for $\mathbb{L}_{N/2}^{(i)}(y_1^{N/2}, \hat{u}_{1,o}^{2i-2} \oplus \hat{u}_{1,e}^{2i-2})$ and $\mathbb{L}_{N/2}^{(i)}(y_{N/2+1}^N, \hat{u}_{1,e}^{2i-2})$, another input $\hat{u}_{2i-1}$ is also required by Type I PE to process the computation. Moreover, for efficient execution of each Type I PE, the value of $\hat{u}_{2i-1}$ needs to be provided on the fly. However, even for the 8-bit decoder illustrated in Fig. 1, the complicated interleaving of odd and even indices makes the straightforward calculation of $\hat{u}_{2i-1}$ inconvenient. In order to solve this inherent problem, the input generating circuit (IGC) for Type I PEs is proposed in this section. Careful investigation has shown that it is possible to generate the required $\hat{u}_{2i-1}$ using the real FFT-like signal flow. For an instance, all the extra input values $\hat{u}_{2i-1}$ for 8-bit polar code decoder can be easily generated according to the following in-place procedure.

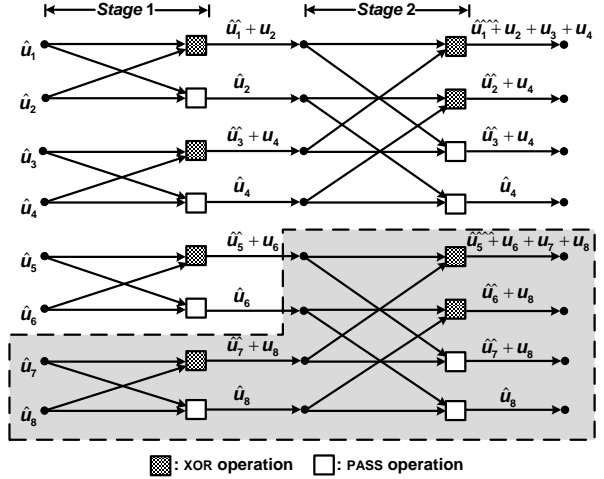

Figure 11: Flow graph of IGC for 8-point polar decoder.

Here, the pass operation process element only lets the lower input get through.

According to Fig. 11, the flow graph can be further simplified with the properties explained as follows:

> 1. The first simplification is to consider that all outputs which are associated with inputs $\hat{u}_{N-1}^N$ are not necessary. Consequently, the shaded region can be removed. Similar concept can be applied to the general case of $N$ inputs. For any *Stage i*, its lower region which contains $2^i$ process elements (PEs), can be removed. For example, the lower 2 PEs of *Stage* 1 and the lower 4 PEs of *Stage* 2 are removed from the flow graph in Fig. 11. Therefore, only $(N/2)(\log_2 N\text{-}1)$ outputs need to be computed.
>
> 2. The second simplification refers to the fact that the PASS operation element □ can be replaced by the wire connection while the flow graph stays functionally the same. According to Fig. 11, element □ only allow the lower input through to the next stage. Thus, if the upper token is not treated as an input to element □ any more, one simple wire which connects the lower token and the output can be employed instead. Meanwhile, complexity of the flow graph can be halved with respect to the former one.

The resulted simplified data flow graph is given in Fig. 12. Both properties have been fully utilized. It is easy to see that totally $\log_2 N\text{-}1$ stages are required and the number of XOR operations is calculated below:

$$[N(\log_2 N - 1) - 2(1 - N/2)/(1 - 2)]/2 \\ = N(\log_2 N - 2)/2 + 1. \quad (14)$$

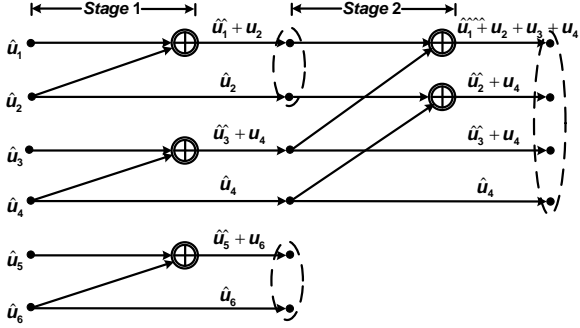

Figure 12: Simplified flow graph of the proposed IGC.

In addition, it is worthwhile to note that with the help of *generator matrix* $G_N$ [1], the same simplified flow graph can be obtained as well. We define $\otimes$ to be the matrix Kronecker product and $n$ equals $\log_2 N$. Then the *generator matrix* $G_N$ is given by the following equation:

$$G_N = B_N F^{\otimes n} = F^{\otimes n} B_N, \quad (15)$$

where $B_N$ is the bit-reversal permutation matrix and $F$ matrix is defined as:

$$F \triangleq \begin{bmatrix} 1 & 0 \\ 1 & 1 \end{bmatrix}. \quad (16)$$

The pipelined architecture of the simplified flow graph illustrated in Fig. 12 can be implemented with the following feed-forward architecture, where totally two XOR-pass elements are required.

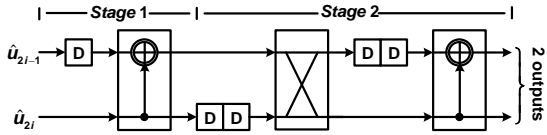

Figure 13: Pipelined feed-forward architecture for 8-bit IGC.

Easily to observe that the proposed pipelined architecture is only suitable for serial inputs. However, as indicated by the time chart illustrated in Fig. 2 (b), every two decoded bits are output in the same clock cycle. In order to work compatibly with this schedule, the architecture needs to be modified to make itself a 2-level parallel processing structure, which is shown in Fig. 14. The control bit $c_1$ is changed with the clock flipping manner and its initial value is set to be 0. By unfolding the $i$-th stage with the *unfolding factor* of $2^{i-1}$, the updated input generating circuit is able to create $2^i$ outputs at the same time, which the original circuit does in $2^{i-1}$ consecutive clock cycles. For example, the unfolded version of the circuit shown in Fig. 13 is given as follows, where $U_i$ denotes the unfolded pipelined architecture which is consists of $i$ stage(s):

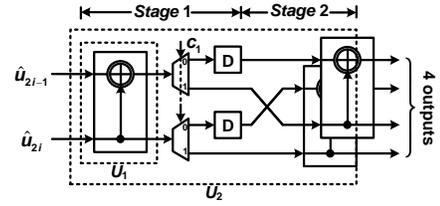

Figure 14: Unfolded version of pipelined architecture in Fig. 13.

In general, for *N*-bit length decoder, since the data structures of IGC are defined recursively for powers of 2, the unfolded pipelined architecture can be constructed with the recurrence relationship. The recursion for the general case is shown in Fig. 15, where module $U_n$ can be constructed based on module $U_{n-1}$ and $N/4$ extra XOR-pass elements. Here, we have $n = \log_2 N - 1$. And the control bit $c_n$ can be obtained by down sampling $c_1$ by a factor of $n$.

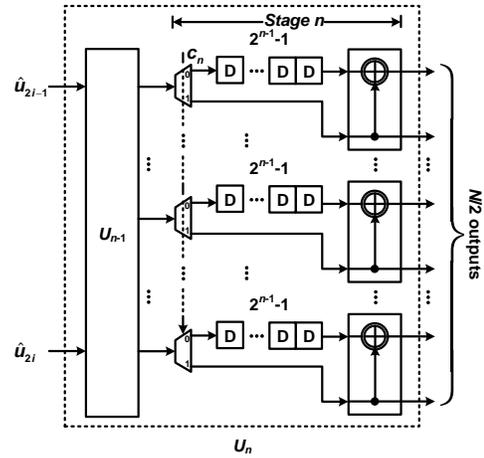

Figure 15: Recursive construction of $U_n$ based on $U_{n-1}$.

Therefore, the total number of XOR-pass elements can be calculated as:

$$\sum_{i=0}^{n-2} 2^i = N/2 - 1. \quad (17)$$

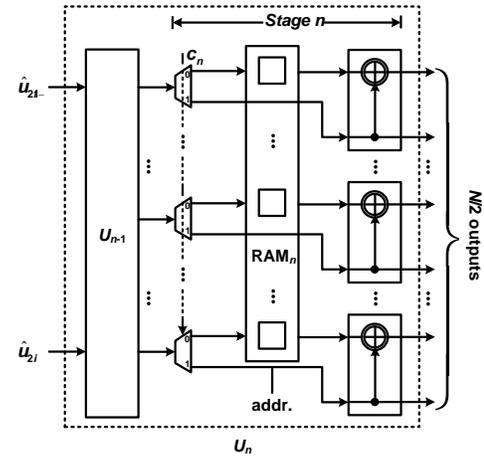

Figure 16: Recursive construction of $U_n$ based on $U_{n-1}$ using RAMs.

Also it is easy to see that the same amount of demultiplexers is required. Finally, as what we expect, the obtained pipelined structure works best with the proposed time chart, which enables all intermediate results to be generated in place without any extra clock cycles.

However, it can be noticed that for *Stage i*, the number of corresponding registers increases with the complexity of $2^i$, which is obviously impractical for polar codes of length over $2^{10}$. One possible approach is to employ memory banks instead of flip-flops, which is shown in Fig. 16. For RAM$_i$, totally $2^{i-1}$ memory elements are required. And the data lifetime is $2^{i-2}-1$ clock cycles, according to which the memory enable signals can be therefore determined.

### E. Pipelined Architecture of the Look-Ahead Decoder

Taking the advantage of the pre-stated modules, the overall pipelined architecture of the proposed look-ahead decoder can be designed accordingly. Without of loss of generality, here we employ an 8-bit polar decoder as an example. Fig.17 shows the proposed pipelined architecture for a 8-bit look-ahead polar decoder, which is composed of the main computation structure and the IGC part. It is a feed-forward pipelined structure that tries to maximize the use of the hardware and minimize the latency of decoding.

decoder architectures, the hardware utilization of each active stage is 100%, other stages still remain idle at the same time. Moreover, for the proposed 8-bit polar decoder, totally 7 clock cycles are required before the next codeword can be processed. Generally, for an *N*-bit polar decoder each codeword needs *N*-1 clock cycles to be properly decoded with the given approach, during which no new codeword could be input to the decoder. Therefore, even only half latency is needed by the proposed scheme, the hardware efficiency remains low for decoders with large *N*. Meanwhile, in synthesizing DSP architectures it is also important to maximize the silicon efficiency of the integrated circuits. One possible approach is to further refine the pipelined architecture which enables new input every clock cycle [6]. According to the decoding time chart depicted in Fig. 2 (b), the new decoding schedule which triples the decoding throughput is given in Table II. In order to process three codewords simultaneously, *Stage* 3 has been duplicated (*Stage* 3') to avoid data contradiction. It is obvious to see that the hardware efficiency for stage *i* (*i* > 1) is 85.7%. And that of *Stage* 1 is 42.9%. Compared with similar approach in [3], the proposed one achieves the same utilization rate at each stage but is better arranged, which can be observed from Table II of [3] easily.

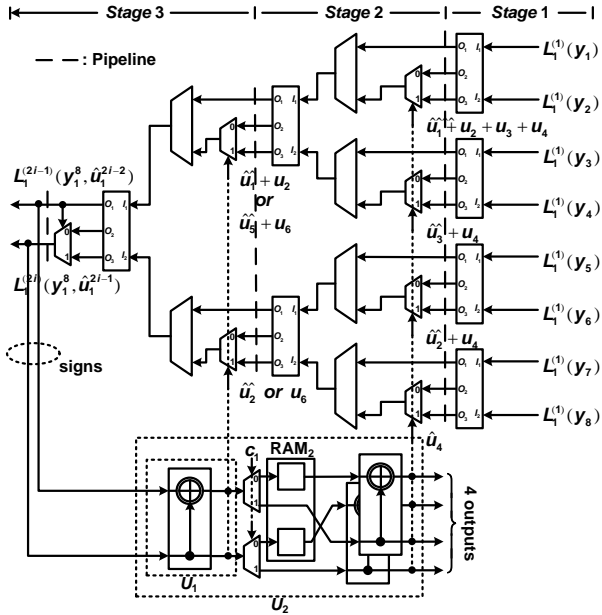

Figure 17:  Pipelined decoder of look-ahead polar codes with *N* = 8.

In general, for the *N*-bit polar decoder, totally *N*-1 incorporated PEs, 2*N*-3 2-to-1 multiplexers, *N*/2-1 XOR-PASS elements, *N*/2-2 1-to-2 demultiplexers, 3(*N*-1) delay elements, and *N*/2-2 bits of RAM are required.

## V. MODIFIED ARCHITECTURES FOR PROPOSED DECODER

In this section, we propose the systematic design methods for three different modified look-ahead polar decoders based on the one illustrated in Fig. 17. DSP-VLSI design techniques are well employed to improve the hardware efficiency while keep the reduced decoding latency unchanged.

*1) Refined Pipelined Architecture of the Look-Ahead Decoder*

It is easy to notice that although for the proposed pipelined

TABLE II  REFINED DECODING SCHEDULES OF 8-POLAR DECODER

| Stage | Clock cycle | | | | | | | |
|---|---|---|---|---|---|---|---|---|
| | 1 | 2 | 3 | 4 | 5 | 6 | 7 | 8 |
| Look-ahead decoding schedule | | | | | | | | |
| 1 | $C_1$ | — | — | — | — | — | — | $C_2$ |
| 2 | — | $C_1$ | — | — | $C_1$ | — | — | — |
| 3 | — | — | $C_1$ | $C_1$ | — | $C_1$ | $C_1$ | — |
| Look-ahead decoding schedule with refined pipelining | | | | | | | | |
| 1 | $C_1$ | $C_2$ | $C_3$ | — | — | — | — | $C_4$ |
| 2 | — | $C_1$ | $C_2$ | $C_3$ | $C_1$ | $C_2$ | $C_3$ | — |
| 3 | — | — | $C_1$ | $C_2$ | $C_3$ | $C_1$ | $C_2$ | $C_3$ |
| 3' | — | — | — | $C_1$ | $C_2$ | $C_3$ | $C_1$ | $C_2$ |

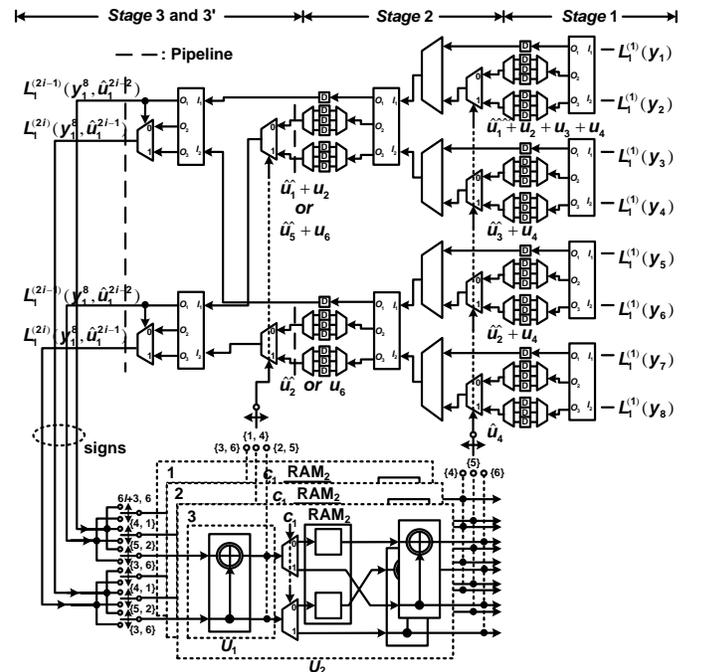

Figure 18:  Refined pipelined decoder of polar codes with *N* = 8.

Without loss of generality, here the authors use the refined architecture of 8-bit polar decoder as an example, which is illustrated in Fig. 18. Other decoders with different code length can be derived accordingly. It is worth noting that *Stage* 3 and 3', which are activated in serial, are generated by unfolding transformation with factor of 2. Keeping in mind that the concurrent 3 inputs are independent, totally 3 copies of IGC are required as a result. According to the recursive construction algorithm of look-ahead time chart given in Section III, it is obvious that for *N*-bit polar decoder the maximum value of concurrent inputs is *N*-1. However, in order to guarantee the non-blocking decoding process, more duplicated stages are required by higher concurrent number *M*. The detailed relationship between concurrent number and hardware consumption can be stated in the following properties.

**Property 1** *For N-bit look-ahead polar decoder, the highest concurrent number M is N-1.*

**Proof** According to recursive algorithm, the decoding time chart will take totally

$$\sum_{i=0}^{\log_2 N-1} 2^i = \frac{2^{\log_2 N}-1}{2-1} = N-1 \quad (18)$$

clock cycles. During the decoding process for a single codeword, *Stage* 1 is only activated in one clock cycle. Therefore, in the rest *N*-2 clock cycles, *Stage* 1 is available for other possible input codewords.∎

**Property 2** *For a given N-bit polar decoder architecture, the $2^i$-1 concurrent version can be derived by duplicating $2^{i-1}$-1 stages, which have the most significant indices, of the $2^{i-1}$-1 concurrent version.*

**Proof** It can be noticed that

$$2^i - 1 = 2(2^{i-1}-1)+1, \quad (19)$$

which is in the same manner that the look-ahead decoding chart is constructed. In order to make 100% hardware utilization of the whole decoder (or certain specific stages), for each decoding stage the number of PEs should stay the same. Since the decoding chart is constructed in the time domain, we only need to apply the same approach in the "stage domain", which results in the method given in Property 2.∎

For example, the 3-concurrent version of 8-bit look-ahead polar decoder in Table II is implemented by adding a duplicated *Stage* 3 to the 1-concurrent version. Moreover, its 7-concurrent version, which can achieve 100% efficiency, can be constructed based on the 3-concurrent version accordingly as follows in Table III. Since

$$\begin{cases} 7=2^3-1 \\ 3=2^{3-1}-1 \end{cases}, \quad (20)$$

the 7-concurrent version can be derived based on the 3-concurrent one by duplicate *Stage* 2, 3, and 3', which have the most significant indices. It can be seen that the proposed 7-concurrent decoder can handle inputs perfectly and achieve 100% utilization rate during *Decoding iteration i* ($i > 1$).

**Property 3** *For any M which satisfies $2^{i-1}-1 < M \leq 2^i-1$, the M-concurrent polar decoder requires the same hardware consumption. And 100% hardware efficiency can be achieved if and only if when $M = 2^i-1$.*

**Proof** According to the proof of Property 2, the proof is immediate and its details are omitted here.∎

**Property 4** *For any M which satisfies $2^{i-1}-1 < M \leq 2^i-1$, the totally number of PEs employed by the M-concurrent polar decoder is $N+2^{i-1}\cdot(i-2)$.*

**Proof** According to Property 1, the number of PEs can be calculated as follows:

$$\sum_{j=i-1}^{\log_2 N-1} 2^j + 2^{i-1}\cdot(i-1)$$
$$= (N-2^{i-1}) + 2^{i-1}\cdot(i-1) \quad (21)$$
$$= N+2^{i-1}\cdot(i-2).∎$$

For example, the 1-, 3-, and 7-concurrent versions of 8-bit look-ahead polar decoder are calculated to have 7, 8, and 12 PEs respectively, which can be easily verified according to Table II and III.

TABLE III 3- AND 7-CONCURRENT DECODING SCHEDULES OF 8-BIT POLAR DECODER

| Stage | Clock cycle | | | | | | | | | | | | | |
|---|---|---|---|---|---|---|---|---|---|---|---|---|---|---|
| | 1 | 2 | 3 | 4 | 5 | 6 | 7 | 8 | 9 | 10 | 11 | 12 | 13 | 14 |
| 3-concurrent look-ahead decoding schedule | | | | | | | | | | | | | | |
| 1 | $C_1$ | $C_2$ | $C_3$ | — | — | — | — | $C_4$ | $C_5$ | $C_6$ | — | — | — | — |
| 2 | — | $C_1$ | $C_2$ | $C_3$ | $C_1$ | $C_2$ | $C_3$ | — | $C_4$ | $C_5$ | $C_6$ | $C_4$ | $C_5$ | $C_6$ |
| 3 | — | — | $C_1$ | $C_2$ | $C_3$ | $C_1$ | $C_2$ | $C_3$ | — | $C_4$ | $C_5$ | $C_6$ | $C_4$ | $C_5$ |
| 3' | — | — | — | $C_1$ | $C_2$ | $C_3$ | $C_1$ | $C_2$ | $C_3$ | — | $C_4$ | $C_5$ | $C_6$ | $C_4$ |
| 7-concurrent look-ahead decoding schedule | | | | | | | | | | | | | | |
| 1 | $C_1$ | $C_2$ | $C_3$ | $C_4$ | $C_5$ | $C_6$ | $C_7$ | ... | | | | | | |
| 2 | — | $C_1$ | $C_2$ | $C_3$ | $C_4$ | $C_5$ | $C_6$ | $C_7$ | ... | | | | | |
| 3 | — | — | $C_1$ | $C_2$ | $C_3$ | $C_4$ | $C_5$ | $C_6$ | $C_7$ | ... | | | | |
| 3' | — | — | — | $C_1$ | $C_2$ | $C_3$ | $C_4$ | $C_5$ | $C_6$ | $C_7$ | ... | | | |
| 2' | — | — | — | — | $C_1$ | $C_2$ | $C_3$ | $C_4$ | $C_5$ | $C_6$ | $C_7$ | ... | | |
| 3'' | — | — | — | — | — | $C_1$ | $C_2$ | $C_3$ | $C_4$ | $C_5$ | $C_6$ | $C_7$ | ... | |
| 3''' | — | — | — | — | — | — | $C_1$ | $C_2$ | $C_3$ | $C_4$ | $C_5$ | $C_6$ | $C_7$ | ... |

## 2) Folded Architectures of the Look-Ahead Decoder

Folding is a technique to reduce the silicon area by time-multiplexing many algorithm operations into single functional unit. However, in most cases the means of folding will result in trading area for time in DSP architectures. Therefore, in order not to affect the pre-stated decoding schedule with more decoding clock cycles, the folding approach employed here is not trivial. For the 8-bit polar decoder, *Stage* 1 is used as a functional unit and all functions of other stages are properly mapped to this stage. Fig. 19 shows the folding transformation result of the circuits illustrated in Fig. 17. The time instances at which the switch executes are also given to better illustrate the operation scheduling of the folded architectures. Apart from 9 switches, which are negligible, the proposed folded architecture only requires the first stage of decoder given in Fig. 17 while keeps the latency-reduced decoding schedule intact. Easy to observe that, for practical applications, 50% of the incorporated PEs can be eliminated as a result. For large *N*, the corresponding hardware efficiency can be as high as twice of that of the non-folded one. Therefore, the folding technique can help to achieve a good tradeoff between the decoding latency and hardware consumption. Admittedly, higher utilization rate can be achieved if the functional unit employs fewer Merged PEs, for which the most extreme consists of only one Merged PE. Though further folding can result in less hardware consumption, the decoding latency will increase drastically due to the required time-multiplexed scheme. Usually, it is recommended to fold the decoder on the base of *Stage* 1 along with $N/3+2$ switches.

and if one design can be pipelined, it can also be rearranged in parallel. Since in all clock cycles except for the first one, less or equal than half of the Merged PEs are active, a 2-parallel architecture is designed here. Take the 8-bit polar decoder as an example, in order to properly process two independent inputs in an interleaved manner, an additional clock cycle is required as follows in Table IV. However, for large *N* the additional clock cycle introduced by the proposed 2-parallel architecture is negligible compared with the long decoding latency. Also, this parallel architecture can achieve as twice throughput as that of the folded one. Note that a duplicated IGC is necessary to satisfy the requirement of two independent inputs.

TABLE IV  NUMBER OF ACTIVE MERGED PEs IN EACH CLOCK CYCLE

| Input | Clock cycle | | | | | | | |
|---|---|---|---|---|---|---|---|---|
| | 1 | 2 | 3 | 4 | 5 | 6 | 7 | 8 |
| $C_1$ | 4 | | 2 | 1 | 1 | 2 | 1 | 1 |
| $C_2$ | | 4 | 2 | 1 | 1 | 2 | 1 | 1 |

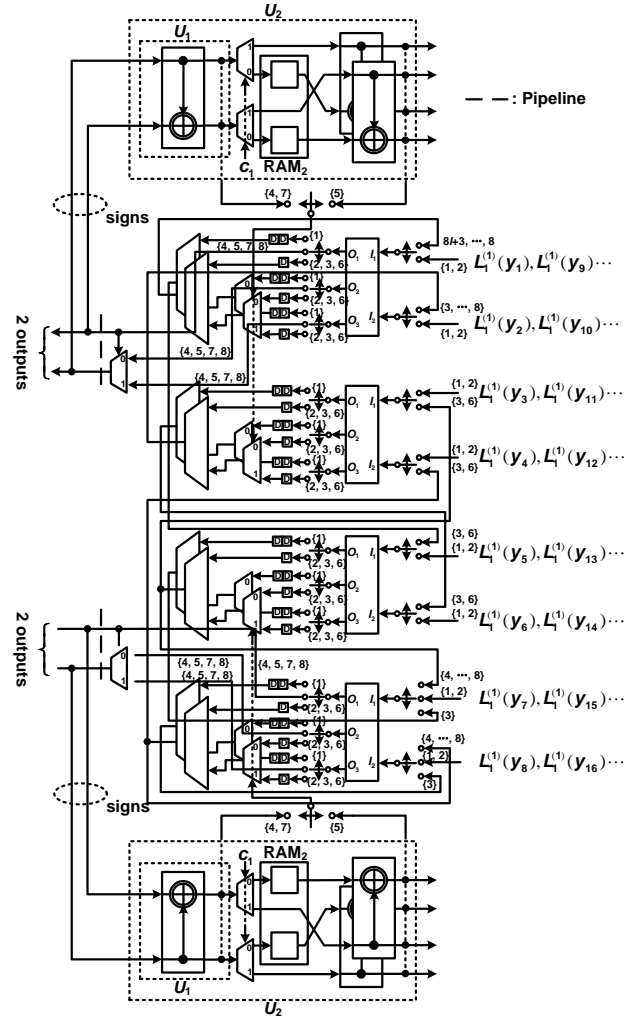

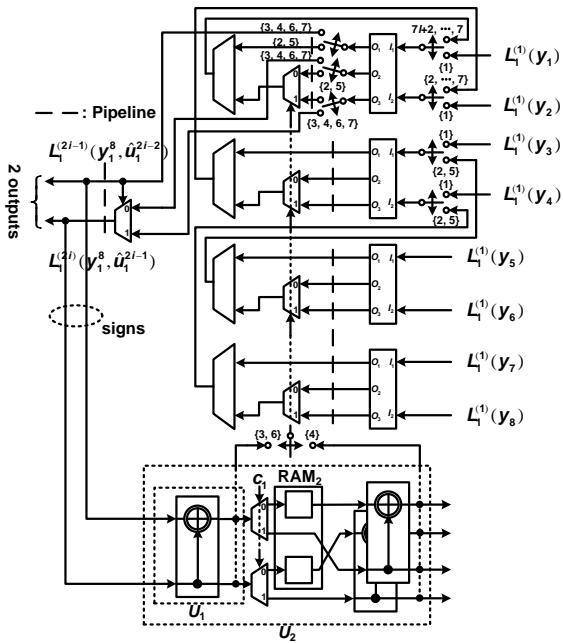

Figure 19: Folded 8-bit polar decoder architectures.

Figure 20: Parallel architectures for 8-bit polar decoder.

### 3) Parallel Architectures of the Look-Ahead Decoder

It is obvious to notice that only during *Clock cycle* 1 can the folded architecture achieves 100% hardware efficiency, which indicates that further improvement is possible. Since parallel processing and pipelining techniques are duels of each other,

On the other hand, for cases in which low hardware consumption is the priority rather than short decoding latency, *L*-parallel processing with $2 < L < N/2$ can be more appreciated. In general, *L*-parallel processing will introduce totally

TABLE V  COMPARISON FOR DIFFERENT POLAR DECODER ARCHITECTURES

| Different designs | | 1st design | 2nd design | 3rd design | 4th design | Tree design[†] | Overlapped design[†] | Line design[†] |
|---|---|---|---|---|---|---|---|---|
| Hardware consumption | | | | | | | | |
| # of Merged PEs | | $N$-1 | $N+2^{i-1}\cdot(i-2)$ | $N/2$ | $N/2$ | $N$-1 | $\sim N+M(\log_2 M/2)/2$ | $N/2$ |
| 1 PE | XOR | $9q$ | | | | | $11q$-3 | |
|  | REG | 0 | | | | | 1 | |
|  | MUX | $6q$ | | | | | $5q$ | |
| # of IGCs | | $N$-1 | $M$ | 1 | 2 | — | — | — |
| 1 IGC | XOR | $N/2$-1 | | | | — | — | — |
|  | RAM | $N/2$-2 | | | | — | — | — |
|  | MUX | $N/2$-2 | | | | — | — | — |
| # of other REGs | | $q(3N$-4) | $(2M+1)q[N+2^{i-1}\cdot(i-2)-i]+2i$ | $q(3N/2+2)$ | $q(9N/2+4)$ | $q(N$-1) | $\sim qM[N+M(\log_2 M/2)/2]$ | $q(N$-1) |
| # of other MUXs | | $q(2N$-3) | $6q[N+2^{i-1}\cdot(i-2)]$ | $q(N$-1) | $q(N+2)$ | 0 | $\sim q[2N+M(\log_2 M/2)]$ | $3q(N/2$-1) |
| Total[‡] | XOR | $\sim 17qN$ | $\sim 21q[N+2^{i-1}\cdot(i-2)]$ | $\sim 17qN/2$ | $\sim 17qN/2$ | $\sim(16q$-3)$N$ | $\sim(18q$-3)$[N+M(\log_2 M/2)/2]$ | $\sim(19q$-3)$N/2$ |
|  | REG | $\sim 3qN$ | $\sim(2M+1)q[N+2^{i-1}\cdot(i-2)]$ | $\sim 3qN/2$ | $\sim 9qN/2$ | $\sim(q+1)N$ | $\sim(M+1)q[N+M(\log_2 M/2)/2]$ | $\sim(q+1/2)N$ |
| Decoding schedule | | | | | | | | |
| Latency | | $N$-1 | $N$-1 | $N$-1 | $N$ | $2(N$-1) | $2(N$-1) | $2(N$-1) |
| Throughput[∗] | | 1 | $M$ | 1 | 2 | 1 | $M$ | 1 |

[†]All designs are proposed by [3].
[‡]Comparison does not include IGC block.
[∗]Normalized results are compared.

$$\sum_{i=1}^{L-1} i = L(L-1)/2 \qquad (22)$$

clock cycles and $L$-1 copies of IGC. In this paper, for consideration of good balance between latency and hardware consumption, only details of the 8-bit 2-parallel decoder are provided in Fig. 20.

## VI. COMPARISON OF LATENCY AND HARDWARE CONSUMPTION

In this section, we compare the decoding latency and the hardware efficiency for the proposed look-ahead polar decoder and its three variants along with state-of-the-art references. Table V lists the comparison results of those designs in terms of both latency, hardware, and some other key metrics such as throughput, efficiency, and so on. The 1st design is the straightforward implementation of the proposed look-ahead decoding schedule, whose counterpart is the *tree design*. The 2nd one is developed based on the first one with refined pipelining and unfolding scheme with counterpart marked as *overlapped design*. The 3rd one is derived by folding the whole architectures of the 1st design to *Stage* 1, whose counterpart is named as *line design*. The 4th design incorporates 2-parallel processing with the 3rd design to achieve even higher efficiency at the expense of an additional decoding clock cycle. Since it is the first approach which succeeds in incorporating both folding and parallel processing together, no counterpart is listed here. For ease of explanation, it is assumed that the $(q, f)$ quantization scheme is employed by all design, where $q$ is the fixed length of LLRs and $f$ is the length of the fraction part.

According to Table V, the most significant point is that benefitting from the proposed look-ahead decoding scheme all the given designs require only half the latency as others do. This advantage makes the proposed decoders more applicable for faster implementations, which always demand code length $N$ higher than $2^{10}$. A good example makes the advantage apparent: no matter what value the concurrent number $M$ is, the 2nd design is able to finish decoding all codewords before its counterpart (*overlapped design*) outputs its first decoded word. Another point is the IGC, which is inspired by the real FFT processor proposed in [4] and can be generated with a nice and easy recurrence relationship, is able to output all control bits required by the multiplexers on the fly. Therefore, no additional clock cycles are needed for computation of $\hat{u}_{2i-1}$, which preserves the advantage of short latency. To the best knowledge of the authors, this is the first detailed design of similar module with such features.

Note from Table V, the authors succeed in giving the detailed architecture for each sub-block. Meanwhile, since reference [3] failed to provide details of the $\hat{u}_s$ computation block, which is the counterpart of the proposed IGC, only the comparison of hardware consumption for the rest blocks is conducted. And the assumption that each 1-bit 2-to-1 multiplexer requires the same silicon area as an XOR gate is used here for estimation [7]. Despite of larger number of registers, which are inherently resulted from the look-ahead scheme, the proposed designs need similar hardware area (in the form of XOR gates' number) compared with their counterparts. The 1st and 2nd designs need 6.25% and 14.27% more hardware than their counterparts, respectively. On the other hand, the 3rd and 4th designs only require 89.47% hardware than the *line design*. And the 4th one managed to achieve half decoding latency and twice data throughput compared with the *line design*. Among the four designs proposed in the manuscript, the 2nd one can achieve the highest throughput and hardware efficiency (100%). The 3rd and 4th decoders require the least number of Merged PEs. And the 4th one can reach the optimal compromise between decoding throughput and area, whose throughput-to-area ratio is twice that of the 3rd one. Those proposed designs make it available to meet demands of different application situations.

## VII. Conclusion

Based on the recursive construction method of decoding time chart, a novel look-ahead SC decoding schedule for polar codes is proposed in this paper, which can halve the decoding latency required by conventional approaches. For efficient hardware implementation issue, a Merged PE is presented by using sub structure sharing technique. Its control signal $\hat{u}_{2i-1}$ can be generated with a real FFT-like diagram. This unique feature is directly applied to develop an efficient input generating circuit (IGC), which works best with the latency-reduced polar decoder architecture. The methodology for designing four different look-ahead polar decoder architectures is presented along with gate-level details. Comparison results have shown that aside from the IGC module, the proposed designs show comparable hardware efficiency while have much shorter decoding latency than their conventional counterparts.